# EXCITONS AND EXCITONIC MOLECULES IN MIXED $Zn(P_{1-x}As_x)_2$ CRYSTALS

O.A.Yeshchenko, M.M.Biliy, Z.Z.Yanchuk

*Physics Faculty, Kyiv Taras Shevchenko University*
*6 Akademik Glushkov prosp., 03127 Kyiv, Ukraine*
*E-mail: yes@mail.univ.kiev.ua*

Low-temperature (1.8 K) excitonic absorption, reflection and photoluminescence spectra of mixed $Zn(P_{1-x}As_x)_2$ crystals were studied at $x = 0.01$, 0.02, 0.03 and 0.05. Energy gap and rydbergs of excitonic B, C and A-series decrease monotonically at the increase of $x$. Spectral half-widths of absorption $n = 1$ lines of B and A-series increase monotonically at the increase of $x$ due to fluctuations of crystal potential. Emission lines of excitonic molecules were observed in photoluminescence spectra of $Zn(P_{1-x}As_x)_2$ crystals. Binding energy of molecule increases at the increase of $x$ that is due to the decrease of the electron-hole mass ratio.

## Introduction

*β-$ZnP_2$* (further $ZnP_2$) and *$ZnAs_2$* crystals are the strongly anisotropic direct-gap semiconductors, which are characterized by the same symmetry group $C_{2h}^5$ (monoclinic syngony). The energy gaps of these crystals are the following: 1.6026 eV for *$ZnP_2$* [1] and 1.052 eV for *$ZnAs_2$* [2,3]. Besides symmetry of lattice, the similarity of $ZnP_2$ and *$ZnAs_2$* takes place in the structure of energy bands and exciton states as well. Namely, three excitonic series are observed in the absorption spectra of these crystals: dipolly allowed C-series at $\mathbf{E} \| Z(\mathbf{c})$, forbidden B-series at $\mathbf{E} \perp Z(\mathbf{c})$ and partially allowed A-series at $\mathbf{E} \| X$ polarizations (see e.g. [1,4-6] for *$ZnP_2$* and [2,3,7] for *$ZnAs_2$*). In the photoluminescence (PL) spectra of these crystals at $\mathbf{E} \| Z(\mathbf{c})$ polarization a series of lines caused by the radiative transitions from the ground $n = 1$ and excited $n = 2,3$ states of allowed S-paraexciton (C-series) is observed (see e.g. [5] for *β-$ZnP_2$* and [2] for *$ZnAs_2$*). Besides this emission series, in the PL spectra of *$ZnP_2$* at $\mathbf{E} \perp Z(\mathbf{c})$ the so-called B-line is observed. Emission B-line is due to the radiative transitions from the ground state of forbidden S-orthoexciton and corresponds to $B_1$-line of absorption B-series. Serial laws of excitons in $ZnP_2$ are characterized by the following parameters [1]: for C-series rydberg is $Ry_C = 41.3$ meV, short-range correction to the hydrogenlike $E(n) \propto n^{-2}$ law is $\Delta_C = 1.4$ meV; for A-series $Ry_A = 41.3$ meV, $\Delta_A = 14.1$ meV; for B-series $Ry_B = 45.8$ meV, $\Delta_B = 0$.[1] Thus, for C and B-series the values of corrections are small with respect to rydbergs, and therefore these series can be fitted with high precision by the law $E(n) = E_g - Ry / n^2$, that gives the following rydbergs values: $Ry_C = 39.6$ meV and $Ry_B = 44.0$ meV.[2] For A-series the correction $\Delta$ is large, and therefore the fitting of this series by above hydrogenlike function gives $Ry_A = 26.5$ meV that differs appreciably from the above value. In $ZnAs_2$ crystals the exciton parameters are the following: $Ry_C = 12.5$ meV and $Ry_B = 15.1$ meV [2,3,7]. Concerning A-series, till present moment an ambiguity exists with respect to identification of some lines of this series with concrete values of the quantum number $n$. It is not known whether there are observed lines $n = 1,2,3$ in absorption spectra or only $n = 1,2$ ones [3]. Therefore, we do not give here the value of A-series rydberg in $ZnAs_2$.

[1] It should be noted that $\Delta_B = 0$ only for B-series lines with $n \geq 2$. The $n = 1$ line deviates from the "pure" hydrogenlike law. The value of B-series rydberg $Ry_B = 45.8$ meV is obtained from the fitting of B-series lines with $n \geq 2$.

In $ZnP_2$ crystals excitonic molecule (EM or biexciton) with rather high binding energy $E^b_{bex} = 6.7$ meV $= 0.15 E^b_{ex}$ exists, where $E^b_{ex}$ is the binding energy of lowest-energy exciton in a state with $n = 1$, to which the B-line in absorption and PL spectra corresponds. Proceeding from the theory [8], high relative binding energy of biexciton $E^b_{bex} / E^b_{ex} = 0.15$ in $ZnP_2$ is explained by the rather small ratio of effective masses of electron and hole $\sigma = m_e / m_h = 0.06$. In $ZnP_2$ EMs manifest themselves in PL spectra due to two-electron and two-photon radiative transitions from the ground state of a molecule [9-11]. Two-electron transitions occur from the ground state of biexciton to states of excitons of C and A-series (further C and A-excitons correspondingly). The result of such transitions is appearance of inverse hydrogenlike series of luminescence (further $M_C$ and $M_A$-series correspondingly). The experimental evidences of existence of the biexcitons in $ZnAs_2$ are absent at present, though on theoretical estimations they should have binding energy $\approx 1.6$ meV.

In present work an object of study was the mixed crystals (solid solutions) of isovalent substitution $Zn(P_{1-x}As_x)_2$, which, as we know, were not studied yet earlier. The mixed crystals of $A^{II}B^{V}$ type, to which $Zn(P_{1-x}As_x)_2$ crystals belong, are poorly investigated. At present the mixed crystals of $A^{III}B^{V}$ type are most investigated [12] (review). For the first time direct exciton transitions in absorption spectra of $A^{III}B^{V}$ crystals were observed in $In_{1-x}Ga_xP$ [13], and also - in $Al_xGa_{1-x}As$ [14]. Later narrow excitonic peaks $n = 1,2$ were observed in absorption spectra of direct-gap $GaAs_{1-x}P_x$ [15]. Also, excitonic states are rather well investigated in $A^{II}B^{VI}$ mixed crystals, e.g. in $CdS_{1-x}Se_x$ [16], $ZnSe_{1-x}Te_x$ [17] etc. In present work the low-temperature (1.8 K) absorption, reflection and photoluminescence spectra of $Zn(P_{1-x}As_x)_2$ crystals were studied at small levels of substitution of $P$ by $As$: $x \le 0.05$. Experimental dependences of parameters of excitons and excitonic molecules on $x$ were obtained.

## Technology of growing of mixed $Zn(P_{1-x}As_x)_2$ crystals

The technological operations of synthesis and growing of $ZnP_2$ crystals were carried out according to [6]. To obtain $Zn(P_{1-x}As_x)_2$ crystals the doping by arsenic was carried out during growth of crystals from a gas phase. Using the previously synthesized and twice purified by resublimation $ZnP_2$ (weight 8 gr.) five portions approximately of identical mass were prepared. Each portion was loaded in quartz ampoule (length - 165 mm, bore diameter - 18 mm) with a cone-shaped spout. As on air the arsenic rather promptly oxidizes, for doping we have utillized synthesized by us compound of arsenic $ZnAs_2$, which at temperature of growth of $ZnP_2$ crystals ($940^\circ C$) dissociates to $Zn$ and $As$. The mass of portion of $ZnAs_2$ was taken so that at given mass of $ZnP_2$ contents of arsenic made a part necessary for us, in relation to contents of phosphorus (in atomic %). One ampoule was control, i.e. $ZnAs_2$ was not loaded there. During pumping-out (to pressure $2\times10^{-5}$ mm of a hg.) the ampoules were heated up to temperature $250^\circ C$ during 1 hour and were unsoldered from vacuum installation by the oxygen burner.

---

[2] This value is obtained from the fitting of the all B-series lines including $n = 1$ one.

## Exciton spectra of $Zn(P_{1-x}As_x)_2$ crystals

Absorption, reflection and luminescence spectra of $Zn(P_{1-x}As_x)_2$ crystals at $x = 0.01$, 0.02, 0.03 and 0.05 and also spectra of pure $ZnP_2$ ($x = 0$) are presented on figs. 1-3. The $Zn(P_{1-x}As_x)_2$ crystals are direct-gap semiconductors as well as $ZnP_2$. One can see from figures that in mixed crystals the same excitonic C, B and A-series are observed, as in pure $ZnP_2$. It pays an attention on itself the doublet structure of an absorption $n = 1$ line of B-series of a crystal with $x = 0.02$ (fig. 1). Proceeding from intensities and half-widths of the components of this doublet, we have made a conclusion, that narrow high-energy component is $n = 1$ line of B-series. An origin of low-energy component, which is missing for crystals with $x \neq 0.02$, is not clear. Let's note, that in contrast to $ZnP_2$ crystals, where in optical spectra the lines up to $n = 7$ for B-series and to $n = 4$ for A and C-series are observed, in spectra of $Zn(P_{1-x}As_x)_2$ crystals of various thickness excitonic lines with $n = 1,2$ only are observed. Probably, this fact is a result of "blurring" of edges of bands, which takes place owing to fluctuations of crystal potential and, accordingly, fluctuations of energy gap, caused by chaotic distribution of $As$ atoms on sites of crystal lattice at substitution by them of atoms $P$. At increase of concentration $x$ a shift of spectral lines to low-energy side occurs, that is caused by decrease of energy gap. It could be expected, taking into account the fact, that in $ZnAs_2$ energy gap is 0.55 eV smaller, than in $ZnP_2$. The dependence $E_g(x)$ is given in fig.4(a). But besides the trivial decrease of $E_g$ at increase of $x$, there is also decrease of rydbergs of excitonic series (fig.4(b)). Values of $E_g$ and rydbergs were obtained by fitting of excitonic series by simple hydrogenlike dependence: $E(n) = E_g - Ry/n^2$.

There were also studied dependences of the half-widths of absorption $n = 1$ lines of B and A-series. These dependences are given on a fig. 5.[3] It is seen, that half-widths of $B_1$ and $A_1$-lines monotonously increase at increase of $x$. As is known, the increase of half-width of exciton lines is the result of fluctuations of crystal potential and respective fluctuations of $E_g$. The theory of influence of fluctuations of composition $x$ on half-width of exciton absorption lines was developed in [18], where two extreme cases were considered. First takes place, if the effective size of area of crystal potential fluctuation $R_D = \hbar/(2MD)^{1/2}$, where $M$ is the total mass of exciton and $D(x) = W(x) - W(0)$ ($W(x)$ is the half-width of exciton line), is much larger than exciton Bohr radius: $R_D >> a_{ex}$. Such situation, as a rule, takes place, if the effective masses of electron and hole are small and differ between itself a little: $m_e \sim m_h$. In this case $\Delta$ should depend on $x$ as

$$D(x) = 0.08\frac{\alpha^4 M^3 x^2 (1-x)^2}{\hbar^6 N^2} \quad , \tag{1}$$

where $\alpha = dE_g/dx$, $N$ is the concentration of sites of lattice, where the substituting atoms ($As$ in our case) can "sit". Such situation takes place in many semiconductors, in particular in $A^{III}B^{V}$ crystals. The second case takes place at $R_D << a_{ex}$. It takes place at $m_h >> m_e$. In this case dependence $\Delta(x)$ has such character:

$$D(x) = 0.5\alpha\left(\frac{x(1-x)}{Na_{ex}^3}\right)^{1/2}. \tag{2}$$

The experimental dependences $D(x)$ of absorption $B_1$ and $A_1$-lines in $Zn(P_{1-x}As_x)_2$ crystals were fitted by expressions (1) and (2). From a fig. 5 it is seen, that the experimental points are badly fitted both by an expression (1) and (2). It becomes clear if to estimate the effective size of area of fluctuation $R_D$ for different $x$. Estimations give: $R_D = 82A$ at $x = 0.01$, $R_D = 62A$ at $x = 0.02$ and $R_D = 30A$ at $x = 0.05$. In $ZnP_2$ the exciton Bohr radius is $a_{ex} = 15A$. It is seen, that in $Zn(P_{1-x}As_x)_2$: $R_D > a_{ex}$, but the condition $R_D >> a_{ex}$ is not fulfilled, i.e. in studied crystals the intermediate case takes place which is, nevertheless, more close to (1). In $Zn(P_{1-x}As_x)_2$ the condition $R_D >> a_{ex}$ is not fulfilled due to rather large total mass of exciton and appreciable difference of electron and hole masses ($m_h >> m_e$). Therefore, since the intermediate case takes place, the experimental dependences were fitted by the function:

$$D(x) = (1-c)D_1(x) + cD_2(x), \tag{3}$$

which is the superposition of functions $D_1(x)$ of type (1) and $D_2(x)$ of type (2), $c$ is the weighting factor. It is seen from fig. 5, that the function (3) fits experimental dependence $D(x)$ very well, which confirms our guess of an intermediate case. At known $\alpha = dE_g / x$, $M$ and $a_{ex}$, fitting $D(x)$ by function (3) gives the possibility to estimate $N$ and $c$. The fitting of dependences $D(x)$ for absorption $n = 1$ lines of B and A-series gives: $N_B = N_A = 0.026A^{-3}$, $c_B = 0.05$, $c_A = 0.32$. Concentration of sites of lattice, where the substituting atoms can "sit", can be estimated as $N \sim a_{P-P}^{-3} = 0.09A^{-3}$, where $a_{P-P} = 2.20A$ is the length of $P-P$ bond in $ZnP_2$ crystal. It is seen, that values obtained from fitting and estimated from length of $P-P$ bond are in quite good agreement, taking into account estimation character of value $N \sim 0.09A^{-3}$. The fact, that $c_A$ is considerably larger than $c_B$, indicates that in dependence $D(x)$ for $n = 1$ line of A-series the function of type (2) $D_2(x)$ makes a considerably larger contribution, than for $n = 1$ line of B-series. It is clear, as A-exciton has considerably smaller binding energy, than B-exciton, and so has larger Bohr radius. This makes A-exciton closer to a case (2) comparing with B-exciton.

[3] An absence of data on the half-widths of lines for crystal with $x = 0.03$ is due to the fact that sample with $x = 0.03$ was rather thick, and therefore full absorption took place in $B_1$ and $A_1$ lines. So, it was not possible to determine correctly the half-widths of these lines. As has been pointed above, half-width of $B_1$ line for crystal with $x = 0.02$ was determined for the high-energy component of doublet.

## Excitonic molecules in $Zn(P_{1-x}As_x)_2$ crystals

In PL spectra of mixed $Zn(P_{1-x}As_x)_2$ crystals, as well as $ZnP_2$ crystals, there are $M_C$ and $M_A$ lines (fig. 3). These lines are due to radiative transitions from the ground state of excitonic molecule to the ground state of C-exciton ($M_C$-line) and ground state of A-exciton ($M_A$-line). EMs rather well investigated in pure $ZnP_2$. As well as in pure $ZnP_2$, the M-lines demonstrate square-law character of dependence of their intensity on excitation intensity, which confirms their biexcitonic nature. Binding energy of EM can be determined from PL spectrum as: $E_{bex}^b = 2E_{ex} - E_{C1} - \hbar\omega_{M_C}$, where $E_{ex}$ is the energy of lowest exciton ground state, which is $n = 1$ state of B-exciton ($B_1$-line), $E_{C1}$ is the energy of the ground state of allowed C-exciton, $\hbar\omega_{M_C}$ is the energy of a photon of $M_C$-line. Obtained dependence of binding energy of biexciton on $x$ is presented on fig. 6. It is seen, that contrary to rydbergs of excitonic series the binding energy of EM increases at increase of $x$ both by absolute value and in relation to binding energy of lowest B-exciton. Such behaviour of $E_{bex}^b$ at increase of $x$ can be explained in such a way. As is known (e.g. from [8]), the ratio $E_{bex}^b / E_{ex}^b$ is a function of the ratio of effective electron and hole masses $\sigma = m_e / m_h$, and at decrease of $\sigma$ the ratio of binding energies of biexciton and exciton increases. At substitution of atoms of one type by atoms of other type (in our case of $P$ atoms by $As$ atoms), the variation of parameters of energy band structure occurs, in particular the variation of effective masses of carriers. The decrease of exciton rydbergs at increase of $x$ is an evidence of decrease of reduced effective mass of carriers, as dielectric constant should not vary considerably, as $\varepsilon$ in $ZnP_2$ and $ZnAs_2$ crystals has close values. Reduced mass $\mu = m_e m_h / (m_e + m_h) = m_e / (1+\sigma)$ at small $\sigma < 0.1$ is $\mu \approx m_e$. Therefore, decrease of $\mu(x)$ reflects decrease of $m_e(x)$. Such behaviour of $m_e(x)$, i.e. decrease of effective electron mass at decrease of a direct energy gap, is well known, e.g. for $A^{III}B^{V}$ semiconductors [12]. The decrease of $m_e(x)$ can explain decrease of $\sigma$, and so the respective increase of binding energy of biexciton. But how effective mass of a hole behaves? The dependence $m_h(x)$ can be estimated by such a way. Knowing the ratio of the reduced effective masses $\mu(x)/\mu(0) = Ry(x)/Ry(0)$ (this equality is carried out under condition of $\varepsilon = const$) and the ratio $\sigma(x)/\sigma(0)$,[4] one can calculate the ratios $m_e(x)/m_e(0)$ and $m_h(x)/m_h(0)$ by the following expressions:

$$\frac{m_e(x)}{m_e(0)} = \frac{\mu(x)}{\mu(0)} \cdot \frac{1+\sigma(x)}{1+\sigma(0)},$$
$$\frac{m_h(x)}{m_h(0)} = \frac{m_e(x)}{m_e(0)} \cdot \frac{\sigma(0)}{\sigma(x)}. \qquad (4)$$

Results of respective estimations are given on fig. 7. It is seen, that at increase of $x$ the electron mass decreases, and the hole mass increases. Thus, the increase of $m_h$ occurs even faster, than decrease of $m_e$. This fact becomes clear if to take into account, that in $ZnP_2$ conduction band originate mainly from $Zn$ ions, and valence band – from $P$ ions [19]. Therefore, substitution of $P$ atoms by $As$ atoms should first of all has an influence on parameters of a valence band, in particular on hole effective mass. Thus, simultaneous decrease of $m_e$ and increase of $m_h$ result in appreciable decrease of the ratio $\sigma = m_e/m_h$, and therefore to the respective increase of binding energy of excitonic molecule.

## Conclusions

In conclusion, we point out on the following results obtained. In mixed $Zn(P_{1-x}As_x)_2$ crystals at small $x$ in optical spectra the same three excitonic C, B, and A-series are observed, as in $ZnP_2$ crystal. At increase of $x$ the decrease of energy gap and rydbergs of the excitonic series occur. At increase of $x$ the spectral half-widths of excitonic absorption lines increase, that is result of increasing role of the fluctuations of composition, and respectively the fluctuations of crystal potential. In PL spectra the lines of excitonic molecules are observed. It is revealed, that at increase of $x$ the binding energy of EM increases. The increase of binding energy of biexcitons occurs owing to decrease of the ratio of effective masses of electron and hole, which is result of simultaneous decrease of electron mass and increase of hole mass.

## Acknowledgments

This work was partially supported by Fund of Fundamental Researches of the Ministry of Science and Technologies of Ukraine (grant No. 2.4/311).

[4] For $ZnP_2$ $\sigma = 0.06$. For $Zn(P_{1-x}As_x)_2$ the values of $\sigma(x)$ can be estimated from the dependence $[E^b_{bex}/E^b_{ex}](\sigma)$ [8] and the experimental dependence $[E^b_{bex}/E^b_{ex}](x)$. Respective estimations give: $\sigma(0.01) = 0.052$, $\sigma(0.02) = 0.045$, $\sigma(0.03) = 0.037$, $\sigma(0.05) = 0.026$.

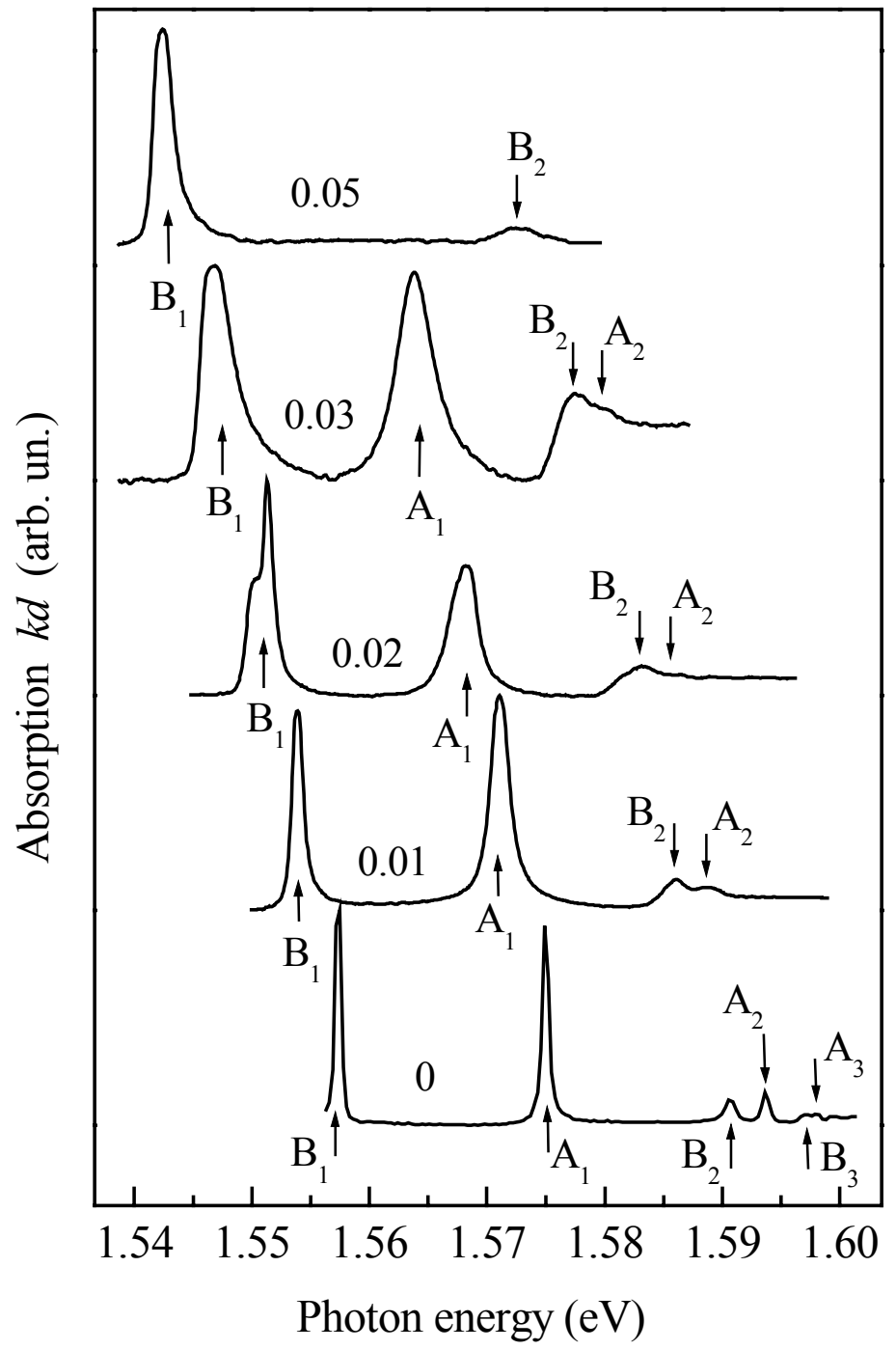

Fig. 1.Absorption spectra of mixed $Zn(P_{1-x}As_x)_2$ crystals at temperature 1.8K. Observation conditions: $\mathbf{q}\perp(110)$, $\mathbf{E}\perp Z(\mathbf{c})$ - for crystals with $x=$ 0, 0.01, 0.02 and 0.03; $\mathbf{q}\perp(100)$, $\mathbf{E}\perp Z(\mathbf{c})$ - for crystals with $x=$ 0.05.

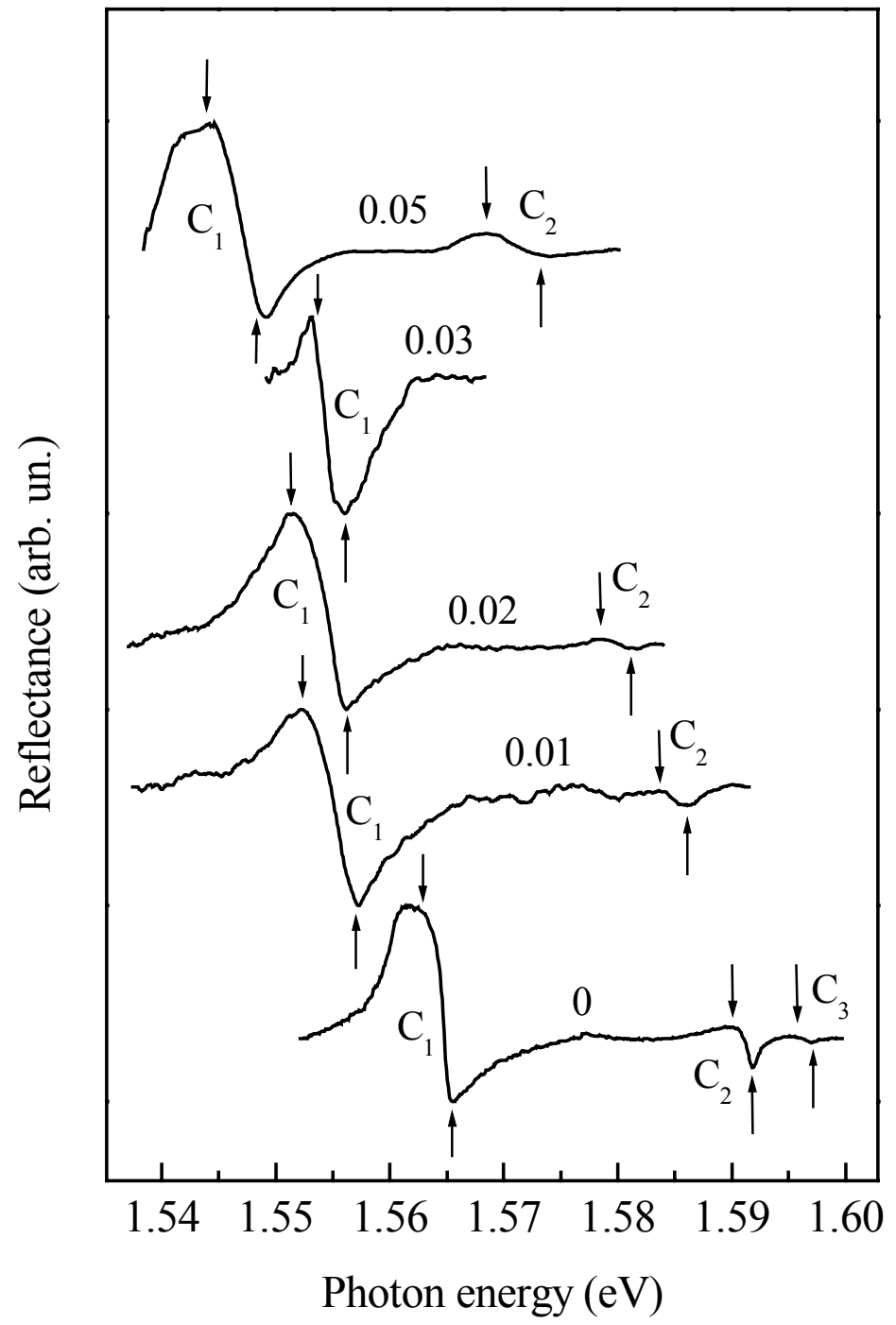

Fig. 2. Reflection spectra of mixed $Zn(P_{1-x}As_x)_2$ crystals at temperature 1.8K. Observation conditions: $\mathbf{q}\perp(100)$, $\mathbf{E}\parallel Z(\mathbf{c})$.

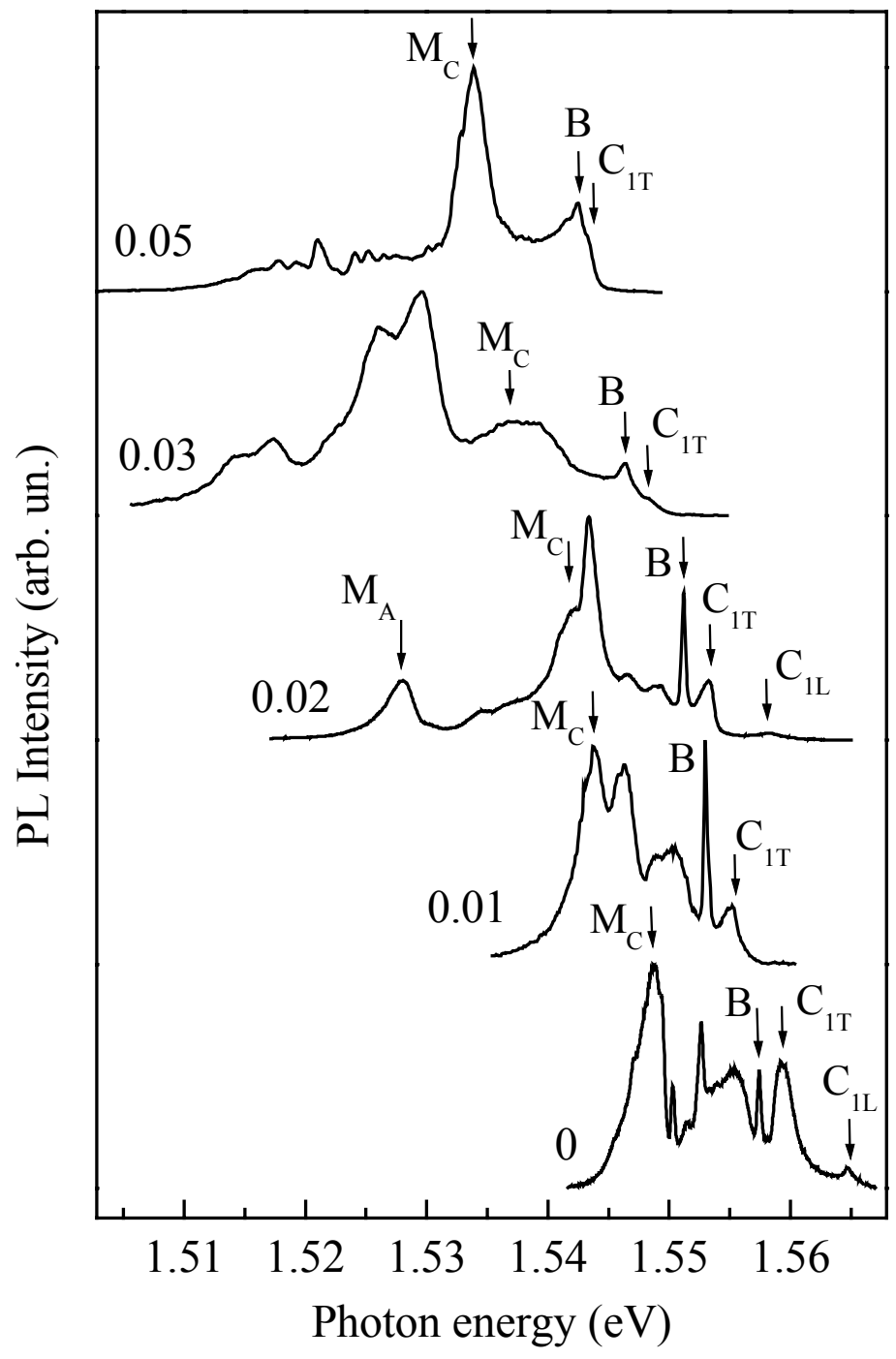

Fig. 3. Photoluminescence spectra of mixed $Zn(P_{1-x}As_x)_2$ crystals at temperature 1.8K.

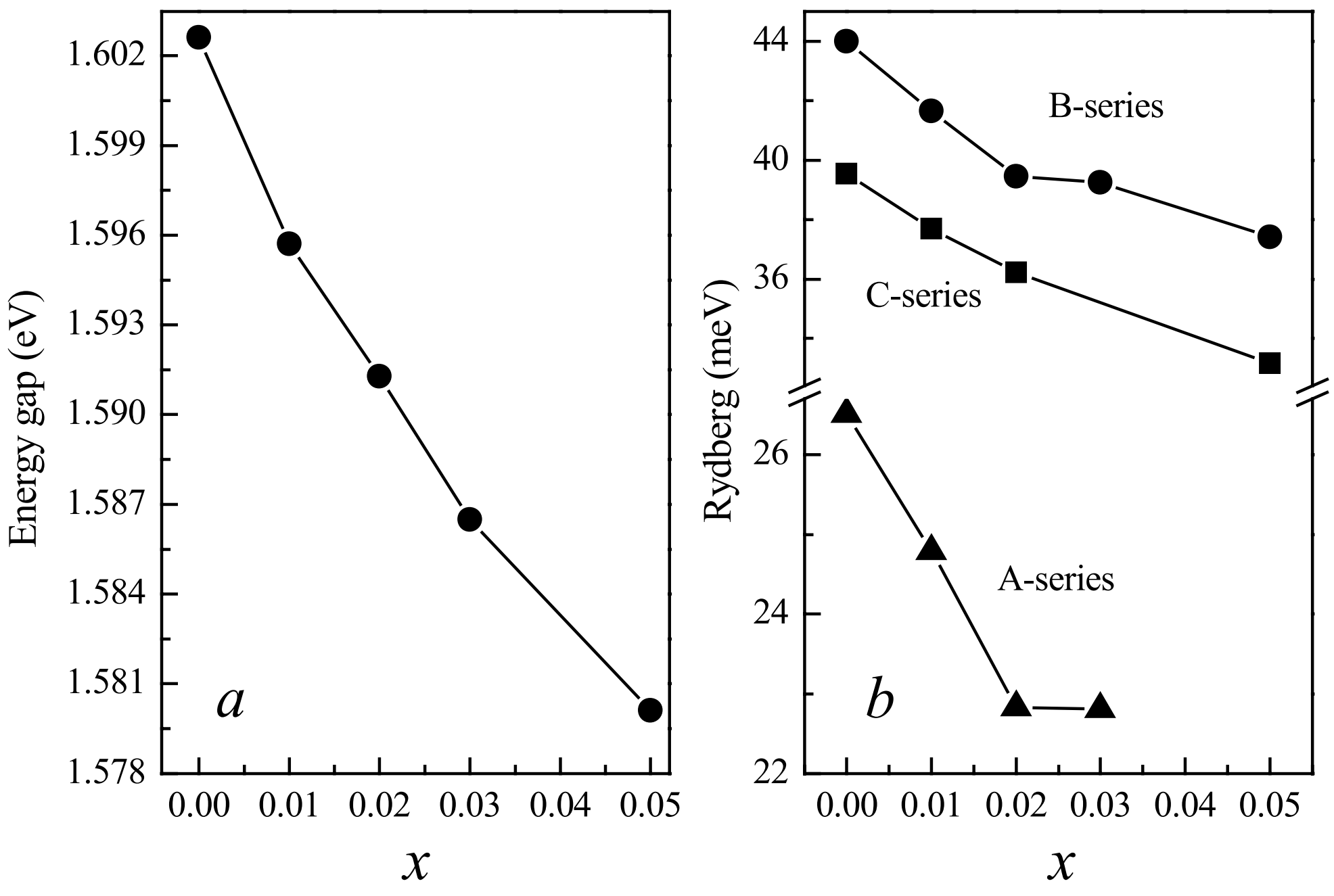

Fig. 4. Dependences of energy gap (a) and rydbergs of excitonic series (b) of $Zn(P_{1-x}As_x)_2$ crystals on level of substitution $x$.

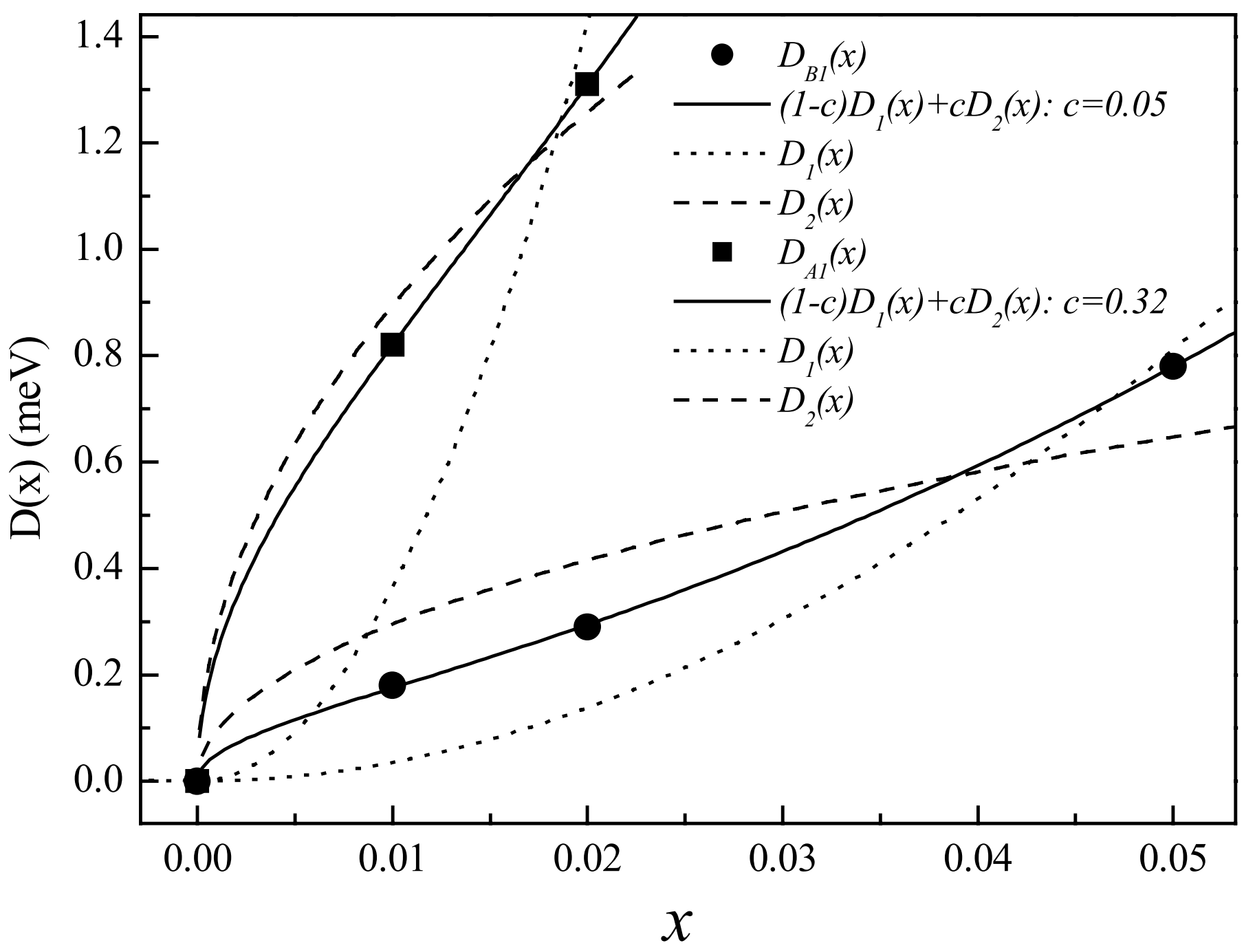

Fig. 5. Dependences of spectral half-widths of absorption $n=1$ lines of excitonic B and A-series of $Zn(P_{1-x}As_x)_2$ crystals on level of substitution $x$.

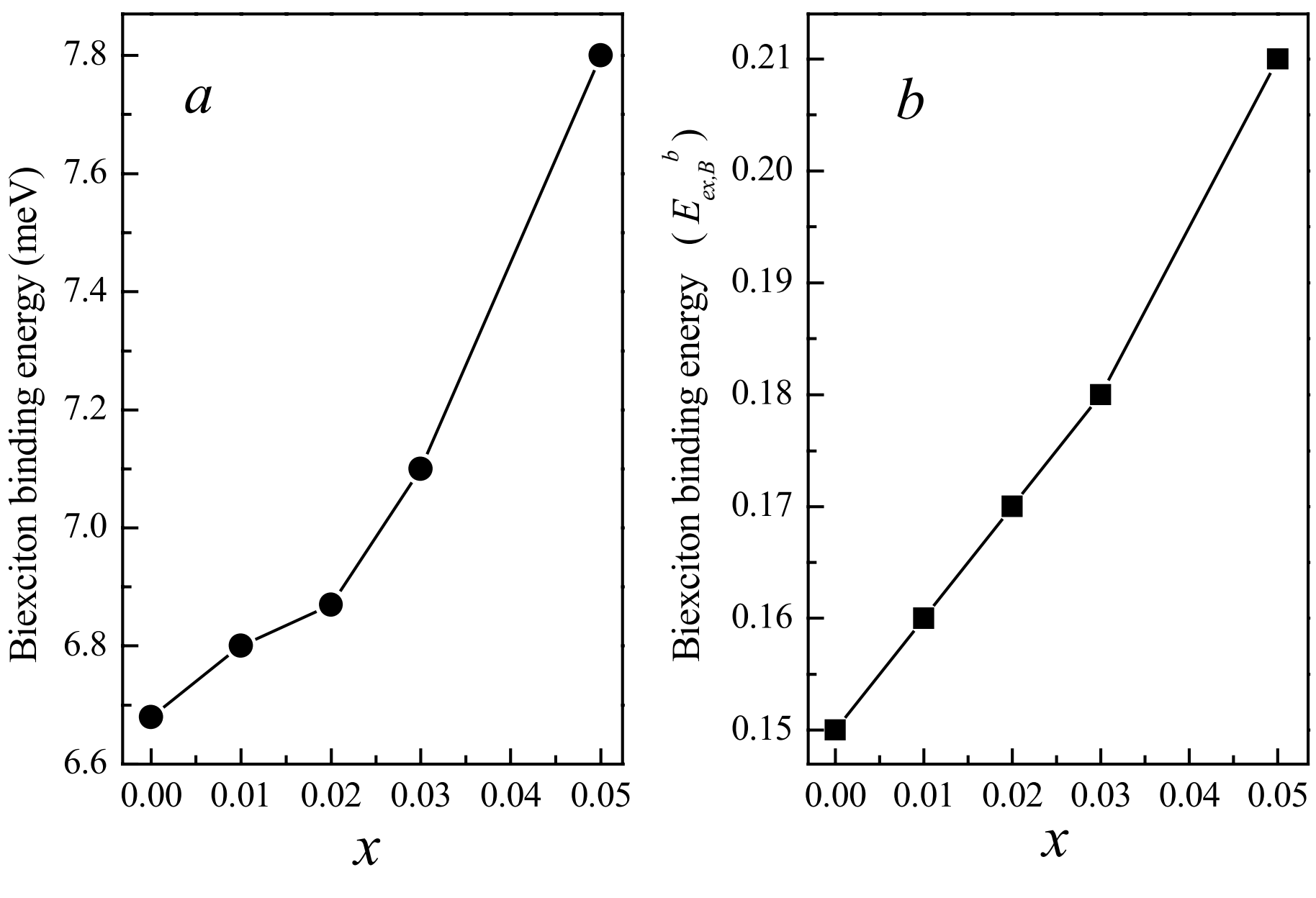

Fig. 6. Dependence of binding energy of excitonic molecule in $Zn(P_{1-x}As_x)_2$ crystals on level of substitution $x$. Binding energy is presented in meV (a), and in units of a binding energy of B-exciton (b).

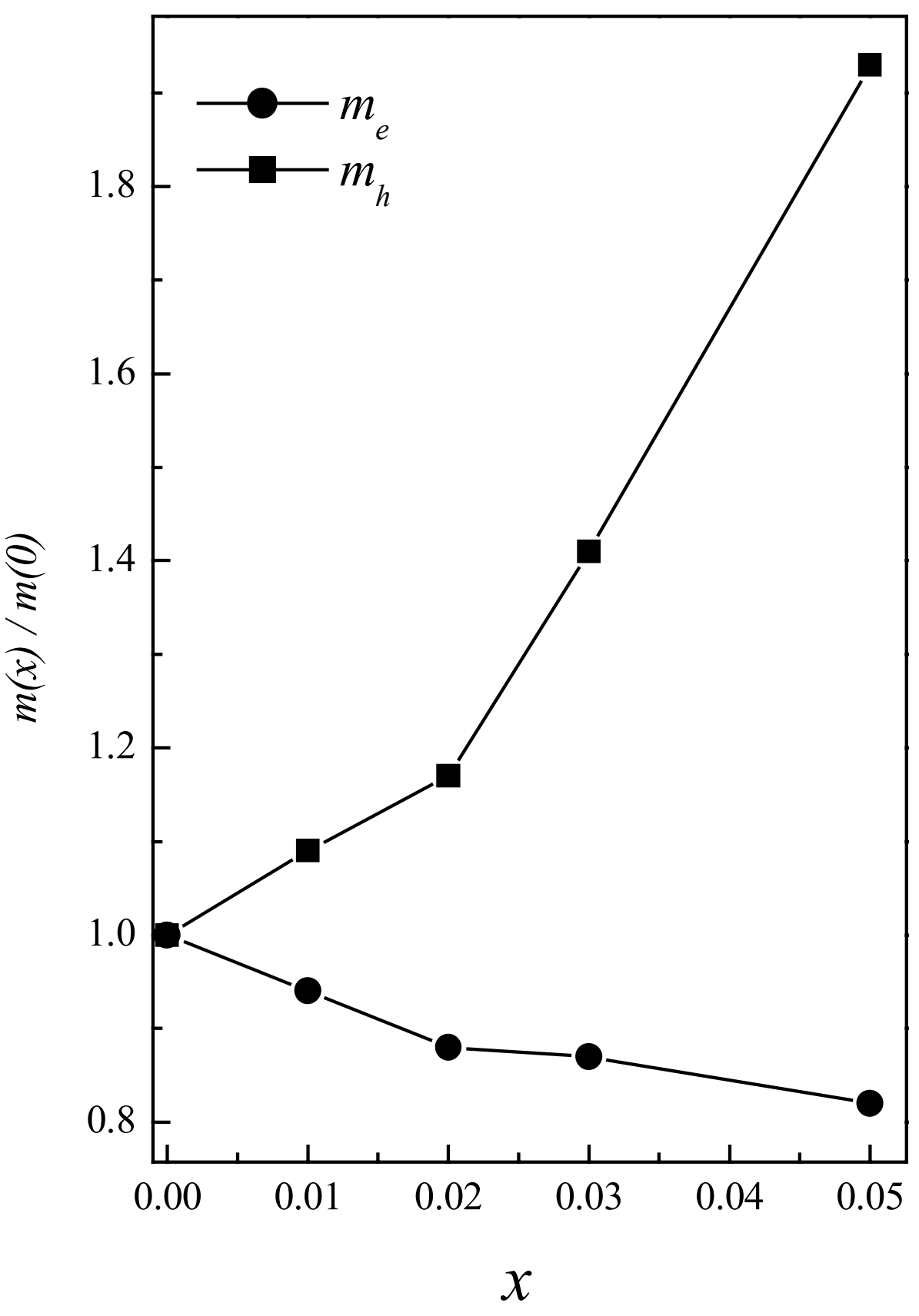

Fig. 7. Calculated dependences of effective masses of electron and hole in $Zn(P_{1-x}As_x)_2$ crystals on level of substitution $x$. $m_{e,h}(0)$ are the masses of carriers in $ZnP_2$ crystal ( $x = 0$ ).